\documentclass[12pt]{article}
\newif\ifdraft
\draftfalse
\ifdraft  \fi
\begin{document}

%
\def\t{{t}}
\def\x{{x}}
\def\v{{v}}
\def\k{{k}}
\def\u{{u}}
\def\n{{n}}
\def\f{{n_0}}
\def\Rset{\hbox{{I\kern -0.2em R}}}
\def\rset{\hbox{{\tiny\rm I\kern -0.2em R}}}
%

\renewcommand{\vec}[1]{\mbox{\boldmath $#1 $}}
\newcommand{\al}{\alpha}
\newcommand{\bt}{\beta}
\newcommand{\gm}{\gamma}
\newcommand{\dl}{\delta}
\newcommand{\ep}{\epsilon}
\newcommand{\varep}{\varepsilon}
\newcommand{\ga}{\gamma}
\newcommand{\lm}{\lambda}
\newcommand{\sg}{\sigma}
\newcommand{\te}{\theta}
\newcommand{\om}{\omega}
\newcommand{\zt}{\zeta}
\newcommand{\la}{\label}

\title{
 Invariance of the relativistic one-particle distribution function.
}
\author{
    { F.~Debbasch }\\
   {\small Universit\'e Paris 6 - C.N.R.S.,}
   {\small L.R.M. (E.R.G.A.),} \\
   {\small Tour 22-12, $4^{\grave{e}me}$ \'{e}tage, bo\^{\i}te 142} \\
   {\small 4 place Jussieu, 75252 Paris Cedex 05, France} \\[1.0em]
  {       J.P.~Rivet} \\
  {\small C.N.R.S., Laboratoire G.D.~Cassini,} 
  {\small Observatoire de Nice,}\\
 {\small F-06304 Nice Cedex 04, France} \\[1em]
{ W.A van~Leeuwen} \\
{\small Instituut voor Theoretische Fysica}\\
{\small Valckenierstraat 65, 1018 XE, Amsterdam, The Netherlands} 
  }

\date{{\sl Physica A}, {\bf 301}, pp.~181--195, 2001}
\maketitle
\ifdraft\newpage\fi
\begin{abstract}
The one-particle distribution function
is of importance both in non-relativistic and relativistic statistical
physics. In the relativistic framework,
 Lorentz invariance is possibly its most fundamental property.
The present article on the subject is a contrastive one\,: we review,
discuss critically, and, when necessary,
complete, the treatments found in the standard literature.
\vspace{0.3cm}
\par\noindent
{ PACS numbers:} 03.30.+p,  05.20.Dd
\par\noindent
{ Keyword:}  distribution function, special relativity.
\end{abstract}

        \section{Introduction}
        \label{sec:Intr}

One of the most important and fruitful concepts in statistical physics
is the concept of phase-space. If one restricts the analysis to 
the non-quantum level, the state of every Galilean system consisting
of $N$ point-like particles can, at any time, be represented by one point in a
$6N$-dimensional phase-space \cite{LL80a}. The statistical behavior of
such a system can then be described by an evolution equation for a distribution 
function, often called phase-space density and 
notated $\rho (t, \vec{r}^N,\vec{p}^N)$, defined, at fixed time $t$,
on this $6N$-dimensional phase-space spanned by the $3N$ positions
$\vec{r}^N = (\vec{r}_1, \ldots, \vec{r}_N)$ and the $3N$ momenta
$\vec{p}^N = (\vec{p}_1, \ldots, \vec{p}_N)$
of the $N$ particles.
In many physically interesting cases, however, the particles which constitute
the system can be considered as weakly interacting only, 
and it then makes sense to introduce the concept of a
one-particle distribution function, often notated
$f(t, \vec{r}, \vec{p})$, and defined ---at any instant $t$--- on a
$6$-dimensional (i.e, the one-particle) phase-space~\cite{H63a}.

In relativistic physics, the concept of a one-particle
distribution function $f$ is also widely used, and it seems even more important
to relativistic statistical mechanics than its Galilean homologue is to
Galilean statistical physics \cite{E71a,I89a}\,: when electromagnetic
interactions are included, it does not seem possible to introduce, at a
relativistic level, an analogue for the Galilean $N$-particle distribution
function $\rho (t, \vec{r}^N, \vec{p}^N)$, since the transmission of
electromagnetic signals can no longer be treated as if they occurred
instantaneously. Therefore, the concept of one-particle distribution has become
one of the cornerstones of non-quantum relativistic statistical
mechanics\,: in practice, the one-particle distribution function
  $f(t, \vec{r}, \vec{p})$ is all one has in relativity.

The natural expression of the particle four-current
in terms of the one-particle distribution function \cite{GLW80a}
strongly suggests that the latter quantity has to be a Lorentz-scalar
for the theory to be consistent within a relativistic framework.
However, the literature on the notion of relativistic
one-particle distribution offers, when submitted to a
critical reading, a rather confusing perspective. Indeed,
various authors differ on the very definition of the concept of distribution
function and, 
consequently, on what should be proved and what has to be put in by hand.
Many authors start from a non-manifestly covariant definition of
the one-particle distribution $f$ that is formally identical
with the usual non-relativistic one, and  then the task remains to show that
such a function is invariant under a change of
reference-frame, \emph{i.e.}, that it is a scalar (see, \emph{e.g.}
\cite{LL75a}, \cite{MTW73a} and \cite{GLW80a}).
To achieve this goal there are, on the one hand,  approaches which are
 a kind of relativistic extensions of the non-relativistic ones.
They, in turn,  fall
into two basically different types\,:
one type based on the so-called invariance of the
volume element in phase-space \cite{LL75a,MTW73a}, the other type based on a 
manifestly covariant rewriting
of the most general  microscopic definition of the one-particle distribution $f$
in terms of mean-values with the help of Gibbs-ensemble averages over
delta functions \cite{GS72a,GLW80a}.
What remains puzzling here is that 
both types of approach have very different physical and mathematical
bases, and do not seem to rely on the same kind of argumentation at all.

On the other hand,  there is a more axiomatic approach to the problem of
introducing a relativistically invariant distribution function $f$; this
other approach
starts from a concept that is
manifestly relativistic and
Lorentz-invariant, namely
the distribution function $f_w$ for the number of particles world-lines
 that cross an arbitrary space-like hyper-surface
in space-time (see, \emph {e.g.}, \cite{E71a} and \cite{I89a}).
The authors who use such a concept \emph{derive} from it
the usual concept of a particle distribution-function and
have then little difficulty in proving that the standard particle
distribution
is also frame-independent. But a direct, microscopic definition of the
distribution function for world-lines, comparable to the standard one for
the particle distribution given in terms of mean-values of delta-functions
over some Gibbs-ensemble, has not yet been given in the literature;
as a consequence it has never been proven that such a world-line distribution
function even exists nor that it is frame-independent. Both assertions
are indeed treated as postulates and this is a rather uncomfortable
situation, especially  considering the fact that other authors, as
mentioned earlier, seem to be able to establish as a theorem the fact that
the particle distribution function is a scalar without having to introduce the
new concept of world-line distribution.  

Our aim is to revisit these issues and to shed some new light on them.
Since the special and general relativistic discussions
exhibit mathematical and physical difficulties which only partly
overlap, we thought it would make things clearer to actually separate
the special and general theory, and to present their treatments in two separate
publications. The present article is, therefore,
devoted to the special relativistic case only, while its following
companion starts where this one stops and addresses the general
relativistic situation.

In this article, we start from the standard definition for the one-particle 
distribution function in phase-space. The crucial issue is then to 
determine whether or not it is possible to establish, by direct reasoning,
that the one-particle-distribution function, so defined, is frame-independent. 

In section~\ref{sec:Attempts}, we review the first type of proof,
based on the so-called invariance of the one-particle  phase-space volume
under Lorentz-transformation \cite{LL75a,MTW73a}.
We show in a mathematically rigorous manner 
that the phase-space volume is \emph{not} Lorentz-invariant; we also
explain why this does not contradict the fact that the one-particle
distribution in phase-space may be Lorentz-invariant and that the 
whole argument is just inconclusive.
In section~\ref{sec:macroh},
we analyze the  proof 
originally developed by de~Groot and Suttorp \cite{GS72a}, 
of which a pedagogical presentation can also be found 
in the book by de~Groot, van~Leeuwen 
and van~Weert \cite{GLW80a}. This proof is based on the most general
definition of the distribution function in phase-space. As such, it makes use
of the concept of an ensemble average and presupposes this
procedure to be covariant. We actually show that this procedure is 
not \emph{a priori} covariant because it relies
on the concepts of macro- and microstates, which are shown \emph{not} to 
be Lorentz-invariant. We therefore introduce the new covariant concepts
of macroscopic and microscopic `histories' and define,
in a manifestly Lorentz-invariant way, statistical ensembles.
The average over these new ensembles is \emph{de facto} 
a scalar procedure and it provides a new definition of the 
one-particle distribution function which ensures that this function is
definitely a Lorentz-scalar.  We 
then show that, contrarily to what might have been expected, the average 
over these covariant ensembles actually comes
down to the usual average over states and that the 
usual relativistic one-particle distribution function is 
therefore, indeed, a Lorentz-scalar; this completes the validation
of the proof  of \cite{GLW80a}.

In section~\ref{sec:worldlines}, we discuss the notion of 
one-particle distribution function for particles world-lines crossing an
arbitrary hypersurface in space-time. We  prove that this notion
only makes sense because the particle-distribution in phase-space is
a Lorentz-scalar. \emph{In other words, if one chooses the axiomatic 
approach to relativistic kinetic theory, postulating that the world-line
distribution function exists is tantamount to postulating that
the one-particle distribution function in phase-space is frame-independent.}
Moreover the world-line distribution function 
turns out to be identical with the standard particle distribution.
Finally, in section~\ref{sec:Discu}, we give an overview of our results 
and we discuss them in some detail.

        \section{Earlier attempts to define the one-particle
                 distribution function}
        \label{sec:Attempts}

The usual definition of the one-particle distribution function in special
relativity is not completely satisfactory. Strictly speaking, it is even wrong.
It is one of the purposes of this section to analyze in detail what happens
exactly when one counts particles in different systems of reference, and to
relate our results to the corresponding ones found in the standard literature.

The final and disappointing conclusion of this section will be that the usual
approaches, 
based on the so-called Lorentz-invariance of the phase-space volume-element,
have all failed if one takes them really seriously. It is our goal
to develop a better approach. This is the subject of the next section,
section 3. In order to make the failure of the earlier attempts as clear as
possible we first follow, in the present section, the usual approach as far
as possible. 

Let ${\mathcal R}$ be an arbitrary Lorentz frame, with respect to which
we want to study a gas of particles. Let 
\begin{equation}
dN(t,\vec{x},\vec{p}) \label{defN}
\end{equation}
be the number of particles which, at time $t$, in ${\mathcal R}$,
have positions and
momenta in the intervals $(\vec{x},\vec{x} + d\vec{x})$ and $(\vec{p},\vec{p}
+ d\vec{p})$, respectively. Let us denote the phase space volume elements
corresponding to these intervals by $d^3x$ and $d^3p$.

Now, the one-particle distribution function $f(t,\vec{x},\vec{p})$,
at time $t$, in ${\mathcal R}$, is defined by the relation\,:
\begin{equation}
dN(t,\vec{x},\vec{p}) = f(t,\vec{x},\vec{p}) d^3x d^3p. \label{1}
\end{equation}
Obviously, the distribution function $f$ has the dimensions of a density in
position and momentum space. We now want to prove that the function $f$,
defined via the equation (\ref{1}), is a Lorentz scalar. To that end we
introduce, next to the reference system ${\mathcal R}$,
a new system of reference, ${\mathcal R}'$
which moves with three-velocity $\vec{v}$ with respect to ${\mathcal R}$.
For reasons of
simplicity, we choose $\vec{v}$ parallel to $\vec{p}$, the momentum of the
particles within $d^3x d^3p$ on which we are now focusing our attention.
Furthermore, we choose the $x$-axis of ${\mathcal R}$ and ${\mathcal R}'$
both parallel to $\vec{v}$.

Because changes occur only in the $x$-directions, the $y$- and $z$-components
of position and momentum variables remain unchanged under the Lorentz
transformations relating the reference systems ${\mathcal R}$
and ${\mathcal R}'$.
We have\,:
\begin{eqnarray}    
ct' & = & \gm (v)\left( ct - c^{-1} v x \right) \label{4}\\
             x' & = & \gm (v) \left( x -  v t \right) \label{5} \\
             y' & = & y  \quad , \quad  z' \quad  = \quad z, \label{6}
\end{eqnarray}
with $v= |\vec{v}| $ is the norm of $\vec{v} = (v,0,0)$, and where $\gm$ is
the `dilatation factor'. The latter is defined, for arbitrary $\vec{v}$, by\,:
\begin{equation}
\gm (v) := \frac{1}{\sqrt{1-\vec{v}^2/c^2}}. \label{7}
\end{equation}
>From (\ref{4})-(\ref{6}) we find\,:
\begin{eqnarray}
cdt' &  = & \gm (v) \left( cdt - c^{-1} v dx \right) \label{8}\\
dx' & = & \gm (v) \left( dx -  v dt \right) \label{9}\\
dy' & = & dy \quad , \quad dz' \quad = \quad dz. \label{10}
\end{eqnarray}
Since, by hypothesis, the particles under consideration occupy, in the
reference frame ${\mathcal R}$, a purely spatial element $d^3x$
characterized by\,:
\begin{equation}
t = \mbox{constant}, \label{11}
\end{equation}
we have, in ${\mathcal R}$\,:
\begin{equation}
dt = 0. \label{12}
\end{equation}
Hence, eqs. (\ref{8})-(\ref{9}) reduce to\,:
\begin{equation}
dt' = -\gm (v) c^{-2} v  dx \label{13}
\end{equation}
\begin{equation}
dx' = \gm (v) dx, \label{14}
\end{equation}
{\sl En passant}, we note that equation (\ref{14}) explains the name dilatation
factor for $\gm(v)$. From (\ref{10}) and (\ref{14}) we find\,:
\begin{equation}
d^3x' = \gm (v) d^3x. \label{15}
\end{equation}
We now come to the transformation in momentum space. If we suppose, for a
moment, that $p^0$, on the one hand, and $p^x,p^y,p^z$ on the other hand,
are independent variables, we have\,:
\begin{eqnarray}
{p'}^0 & = & \gm (v) \left( p^0 - c^{-1} v  p^x \right) \la{17}\\
{p'}^x & = & \gm (v) \left( p^x - c^{-1} v  p^0 \right) \la{18}\\
{p'}^y & =  & p^y \quad , \quad p'^z \quad = \quad p^z, \la{19}
\end{eqnarray}
and thus\,:
\begin{eqnarray}
{dp'}^0 & = & \gm (v) \left( dp^0 - c^{-1} v  dp^x \right) \la{20} \\
{dp'}^x & = & \gm (v) \left( dp^x - c^{-1} v  dp^0 \right) \la{21} \\
{dp'}^y & = & dp^y \quad , \quad dp'^z \quad = \quad dp^z. \la{22}
\end{eqnarray}
However, $p^0$ and $\vec{p}$ are not independent. From the normalization
of the four-momentum $p^{\mu} p_{\mu} = m^2c^2$ we find\,:
\begin{equation}
{p^0}^{2} = m^2c^2 + {p^x}^2 + {p^y}^2 + {p^z}^2, \la{23}
\end{equation}
or
\begin{equation}
dp^0 = \frac{1}{p^0}  \left( p^xdp^x + p^y dp^y + p^zdp^z \right). \la{24}
\end{equation}
Substituting (\ref{24}) into (\ref{21}) we obtain\,:
\begin{equation}
{dp'}^x = \gm (v) \left( 1 - \frac{v}{c}
 \frac{p^x}{p^0} \right) dp^x -
c^{-1} v \gm (v) \frac{p^ydp^y+p^zdp^z}{p^0}, \la{25}
\end{equation}
or, equivalently, using (\ref{17})\,:
\begin{equation}
dp'^x = \frac{{p'}^0}{p^0}\ dp^x - \frac{v}{c}\ \gm (v)
\ \frac{p^ydp^y + p^zdp^z}{p^0}. \la{26}
\end{equation}
Hence, using also (\ref{22})\,:
\begin{equation}
{dp'}^x \wedge {dp'}^y = \frac{{p '}^p}{p^0}dp^x \wedge dp^y - \gm (v)
\frac{v}{c}   \frac{p^z}{p^0}\ dp^z \wedge dp^y, \la{27}
\end{equation}
since the term $ {dp'}^y \wedge dp^y = dp^y \wedge dp^y$ cancels. Similarly,
we find\,:
\begin{equation}
{dp'}^x \wedge {dp'}^y \wedge {dp'}^z = \frac{{p'}^0}{p^0}dp^x \wedge dp^y
\wedge dp^z, \la{28}
\end{equation}
or
\begin{equation}
\frac{{dp'}^x \wedge {dp'}^y \wedge {dp'}^z}{{p'}^0} = \frac{dp^x \wedge dp^y
\wedge dp^z}{p^0}, \la{29}
\end{equation}
or, equivalently\,:
\begin{equation}
\frac{d^3p'}{{p'}^0} = \frac{d^3p}{p^0}, \la{30}
\end{equation}
a well-known result.

In short, we find that (\ref{15}) and (\ref{30}) imply\,:
\begin{equation}
d^3x' d^3p' = \gm (v) \frac{{p'}^0}{p^0}\ d^3x\ d^3p. \la{32}
\end{equation}
An alternative form for ${p'}^0$ (\ref{17}) is\,:
\begin{equation}
{p'}^0 = \gm (v)\ p^0 \left( 1 - c^{-1}v   \frac{u}{c} \right). \la{32a}
\end{equation}
where $u$ is the norm of the particle three-velocity
$c\vec{p} / p^0 = (u,0,0)$.  With (\ref{32a}) we find from (\ref{32})\,:
\begin{equation}
d^3x' d^3p' = \frac{1-vu/c^2}{1- v^2/c^2}\ d^3 x d^3 p, \la{33b}
\end{equation}
where we used the definition (\ref{7}) of $\gm (v)$. 
We did not encounter the result (\ref{33b}) in the literature. In the
particular case in which the arbitrary reference frame
 ${\mathcal R}'$ coincides
with the rest- or comoving-frame of the particles
which move with momentum $\vec{p}$ with respect
to ${\mathcal R}$, we have\,:
\begin{equation}
u = v. \la{33c}
\end{equation}
Let us denote this particular system ${\mathcal R}'$ by
${\mathcal R}^{\ast}$, and the position
and momentum of particles in this particular co-moving system of reference
by $x^{\ast}$ and $p^{\ast}$. We then find from (\ref{33b}) and (\ref{33c})\,:
\begin{equation}
d^3x^{\ast} d^3 p^{\ast} = d^3x\ d^3p. \la{38}
\end{equation}
Hence, $d^3x^{\ast} d^3 p^{\ast}$ is a scalar. However, contrarily to general
belief, the phase space element $d^3x d^3p$ is \emph{not} a Lorentz scalar,
as is seen from (\ref{33b}). Let us digress a little bit on this point. 

Equation (\ref{15}) is a result valid for any Lorentz transformation, from
one system of reference to another, arbitrary system of reference. In
particular, we thus have\,:
\begin{equation}
d^3x^{\ast} = \gm (u) d^3x, \la{39}
\end{equation}
where $u$ is the velocity ${\mathcal R}^{\ast}$ with respect to
${\mathcal R}$.  Hence, combining
(\ref{15}) and (\ref{39}), we have\,:
\begin{equation}
d^3x^{\ast} = \frac{\gm (u)}{\gm(v)}\ d^3x'. \la{40}
\end{equation}
This relation is not always found in the existing literature. In the textbook
`{\it The classical Theory of Fields\/}', Landau and Lifschitz claim
(section 10) that\,:
\begin{equation}
d^3x^{\ast} = \gm (u) d^3x', \la{41}
\end{equation}
basing themselves on (\ref{39}) only, and thus forgetting the step leading to
eq.~(\ref{15}). Misner, Thorne and Wheeler, in their textbook
`{\it Gravitation\/}', in Box 22.5, derive (\ref{38}) and then conclude that
the six-dimensional phase space element is invariant, which it is not, as
implied by (\ref{33b}).

The point missed in most treatments encountered in the literature is that,
in any given reference-frame, 
the volume elements which enter definition~(\ref{1}) for the distribution-
function have to be considered at a fixed time in this 
reference-frame. In other words, the
points of the volume element $d^3x$ in some system of reference ${\mathcal R}$
should be points on a hypersurface of the form $t =\hbox{constant}$ in that
four-space ${\mathcal R}$.
In our treatment this is made apparent by 
 eq.~(\ref{11}); but, if this is so in one reference-frame $\mathcal R$,
this is not so in any other reference-frame $\mathcal R'$:
the same points in space-time occupy the space-volume $d^3x'$ in 
${\mathcal R'}$
but do not belong to a hypersurface $t' =\hbox{constant}$,
since $dt' \neq 0$, as follows from eq.~(\ref{13}). It makes therefore
no sense to count these points in $\mathcal R'$ by using the one-particle
distribution-function in $\mathcal R'$, which is {\sl a priori} suitable 
only for counting points on hypersurface of the form $t' =\hbox{constant}$
(see again definition~(\ref{1})).

The preceding considerations are all related to eq.~(\ref{1}), the defining
relation of the one-particle distribution function
 $f(t,\vec{x},\vec{p})$.

Being a number, the left-hand side of (\ref{1}) is a scalar, which can be
calculated in any reference system. This does not imply, however,
that (\ref{1}) can be used as the defining expression of the distribution
function $f(t,\vec{x},\vec{p})$ in an arbitrary system of reference
${\mathcal R}'$. Let us again elaborate on this point with some more detail.

The inverses of the Lorentz transformations (\ref{4})--(\ref{7}) and
(\ref{17})--(\ref{19}) may be used to express $t$, $\vec{x}$ and $\vec{p}$ in\
terms of $t'$, $\vec{x'}$ and $\vec{p'}$.
We may thus introduce $\tilde f$ defined by\,:
\begin{equation}
\tilde{f} (t',\vec{x'},\vec{p'}) :=
f(t(t',\vec{x'}),\vec{x}(t',\vec{x'}),\vec{p}(\vec{p'})). \la{42}
\end{equation}
With (\ref{42}) and (\ref{33b}) we can reexpress (\ref{1}) in the new
coordinates $(t',\vec{x'},\vec{p'})$, which yields\,:
\begin{equation}
f (t,\vec{x},\vec{p})d^3x d^3p = \tilde{f} (t',\vec{x'},\vec{p'})
\frac{1-v^2/c^2}{1-u^2/c^2}\ d^3x' d^3p'. \la{43}
\end{equation}
In the particular frame ${\mathcal R}^{\ast}$ co-moving with the particles with
momentum $\vec{p}$ we have\,:
\begin{equation}
f (t,\vec{x},\vec{p}) d^3x d^3p = \tilde{f}
(t^{\ast},\vec{x}^{\ast},\vec{p}^{\ast}) d^3x^{\ast} d^3p^{\ast}. \la{44}
\end{equation}
Together with (\ref{1}), eq.~(\ref{44}) suggests that
$\tilde{f} (t^{\ast},\vec{x}^{\ast},\vec{p}^{\ast})$ is the one-particle
distribution function. However, this is not so, since the volume element
$d^3x^{\ast}$ in (\ref{44}) should be a part of a hyperplane $t^{\ast} =$
constant, which it is not, since $dt^{\ast} \neq 0$ [cf. eq.~(\ref{13})].

More generally, the combination
$\tilde{f}(t', \vec{x'}, \vec{p'})(1-v^2/c^2)(1-uv/c^2)^{-1}$
is not the one-particle distribution function in ${\mathcal R}'$,
since $d^3x'$
in (\ref{43}) is not a part of the hyperplane $t' =$ constant [again,
see eq.~(\ref{13})].

To summarize\,: the number $dN$ of eq.~(\ref{1}) indeed is a
Lorentz scalar, as is generally stated. It is simply defined, in any
reference system, as the number of particles which, at time $t$ in
${\mathcal R}$,
occupy the phase-space element $d^3x d^3p$ centered around $(\vec{x},\vec{p})$.
This number, of course, can be evaluated in any Lorentz frame. Indeed,
eq.~(\ref{43}) gives it expression in ${\mathcal R}'$.
However, $dN$ \emph{cannot} be
interpreted as a number of particles in ${\mathcal R}'$.
Therefore, it \emph{cannot}
be linked with the one-particle distribution function in that reference system.
Thus the above calculations do not offer any clue as to what the correct
distribution in ${\mathcal R}'$ is, and the usual approach is inconclusive.
This is the
disappointing conclusion referred to in the introduction to the present section.

        \section{The concept of macrohistory}
        \label{sec:macroh}

The only other direct proof that, in quite general a context, the relativistic
one-particle distribution function in phase-space is a Lorentz-scalar, has been
proposed in \cite{GS72a,GLW80a} . We will first review
rapidly the basics
of this proof and then show that, in order for it to be fully consistent
with the principles of Einstein's relativity, one must introduce the new 
concept of `macrohistory' to replace the usual Galilean concept of 
`macrostate'.

    \subsection{A manifestly covariant expression
                    for the distribution function}
    \label{ssec:reviewproofdeG.}

The basic idea behind the proof proposed in \cite{GS72a,GLW80a}
is to find a manifestly
covariant expression for the distribution function in phase-space, without
having to introduce the concept of world-line distribution function. To achieve
this goal, the authors start from the standard, apparently
frame-dependent definition\,:
\begin{equation}
f(t, {\vec x}, \vec{p}) = \Big \langle \sum\limits_{r}{} \delta({\vec x} -
{\vec x}_r(t))\
\delta(\vec{p} - \vec{p}_r(t)) \Big \rangle,
\label{eq:deffgen}
\end{equation}
where the sum extends to all particles in the system and the outer brackets
`indicate an ensemble average'.
In the relativistic framework, the space and
time degrees of freedom are but coordinates in what is called the
four-dimensional space-time. It is therefore quite natural to introduce
in relativistic statistical 
physics an 8-dimensional `extended' one-particle phase-space, where a point
has typically $(t, {\vec x}, p^0, \vec{p})$ as coordinates. In such
a phase-space, $p^0$ is understood as an independent quantity, not 
necessarily related to $\vec{p}$. Relativistic calculations are usually 
carried out more easily in this phase-space than in the traditional, 
`more physical', 6-dimensional one; in the end, physical results can be
recovered by restricting every equation to the mass-shell or, more
precisely, to the sub-manifold of the mass-shell where $p^0 > 0$.
In this spirit, de~Groot, Suttorp, 
van~Leeuwen and van~Weert introduce another function, $\mathcal F$,
defined over the `extended' 8-dimensional one-particle phase-space by\,:
\begin{equation}
\mathcal F(t, \vec{x}, p^0, \vec{p}) = 2 \theta(p^0)\
 \delta(p^2 - m^2c^2)\ f(t, \vec{x}, \vec{p}),
\label{eq:defF}
\end{equation}
and prove that $\mathcal F$ is actually the ensemble average of
a Lorentz-scalar.
More precisely, their basic result is that $\mathcal F$ can be written\,:
\begin{equation}
\mathcal F(t, \vec{x}, \vec{p}) = 
 c \sum_{\omega \in \Omega} w_{\omega} \sum_{i}
 \int {\delta}^{(4)} \left(x - X_{i\omega}(s_{i}) \right)
 {\delta}^{(4)} \left(p - P_{i\omega}(s_{i}) \right) ds_{i},
\label{eq:dRS7}
\end{equation}
where the index $i$ labels the particles in the system as well as their
trajectories $\left(X_{i}(s_{i}), P_{i}(s_{i})\right)$ in the extended
phase-space, each trajectory being parameterized by its
proper-time $s_{i}$. The sum over 
$\omega$ is the mathematical expression for the statistical averaging
and $w_{\omega}$ represents the weight associated to each element
$\omega$ in the statistical ensemble $\Omega$.

Since the product of the theta-function by the delta-distribution,
as it appears on the right-hand
side of (\ref{eq:defF}) is a  Lorentz-scalar, the authors conclude that
the distribution function $f$ itself is Lorentz-scalar.
This conclusion is indeed
warranted, but only if the ensemble averaging procedure in itself does not
change the transformation character (variance) of the quantity
 to which it is applied, 
which is the case if the statistical ensemble $\Omega$ and the 
coefficients $w_{\omega}$ are Lorentz-scalars. We will now show
that this is not \emph{a priori} the case for
ensemble averages defined in the usual, Galilean way because
the notions of macro- or microstate are not themselves covariant;
we will therefore introduce
the new concepts of macro- and microhistories and show with their help
that, contrarily to what might have been expected, the traditional
ensemble average is indeed a covariant operation.

    \subsection{Relativistic covariant ensembles}
    \label{ssec:relatens.}
In Galilean statistical physics, ensembles are defined via the concept of
macrostate. A macrostate of a system is defined by the values taken
by  certain macroscopic quantities or macroscopic fields. The nature and
number of the macroscopic quantities are, 
to a certain extent, arbitrary, although some usual or natural choices exist.
If, for instance, one wants to study a perfect
gas out of equilibrium,  macroscopic quantities often used to 
define a macrostate are the particle density field, the velocity field
and an arbitrary `thermodynamical' field, such as the temperature field.
In some contexts, it seems useful to extend the number of fields,
as is commonly done within the framework of Extended Thermodynamics theories.

To any macrostate of a given system correspond many systems
that are macroscopically the same, but are different on a microscopic scale.
A collection of systems that differ microscopically, but are identical
macroscopically, is what is called an ensemble.

The essential point to realize is that,
in Galilean Physics, the values of the various
macrofields define the macrostate of the system at a given time \emph{in the 
reference frame where the statistical study is carried out}. 
Similarly, microstates in Galilean Physics are always states of the system
at a given time in a chosen reference frame. In the Galilean
world, this poses no problem since time is invariant by a change of
reference frame. But this is not the case anymore in the relativistic framework.
To render this discussion more precise, let 
$A^{\mu...\nu}$ be one of the macrofields, 
the value of which defines a macrostate
of the system under consideration. Classical examples in relativistic
hydrodynamics are the particle current density $j^\mu$,
the entropy flux density $S^\mu$ and the stress-energy tensor $T^{\mu\nu}$.
To begin with, let us choose  to study the system  in a given inertial
frame $\mathcal R$.
The macro-state of the system in $\mathcal R$ at time $t$ is defined
by the values taken by the macrofield $A^{\mu...\nu}$ and all other
macrofields on the hypersurface
of space-time ($t =$ const.). The macrostate of the system, at a fixed time $t$
in $\mathcal R$, is therefore defined by the collection of numbers 
\begin{equation}
\hat{A}^{\mu ...\nu}({\vec x}):=A^{\mu ...\nu}(t, {\vec x}) ,
\label{eq:AAtildeR}
\end{equation}
The change of the macrostate under a Lorentz-transformation
characterized by the tensor ${\Lambda^{\alpha }}_\mu$ can be
investigated by applying
$\Lambda$ directly to (\ref{eq:AAtildeR}). One obtains immediately\,:
\begin{equation}
A'^{\alpha \ldots \beta }(t', {\vec x}') = {\Lambda^{\alpha }}_\mu
\dots {\Lambda^{\beta }}_\nu \hat{A}^{\mu ...\nu}(\vec x),
\label{eq:A'Atilde}
\end{equation}
where $t'$ and ${\vec x}'$ are related to $t$ and $\vec{x}$ by the same Lorentz
transformation. In particular, $t'$ in (\ref{eq:A'Atilde}) depends on both $t$
and $\vec x$ or, equivalently, on $t$ and ${\vec x}'$. The variable
$t'$ is therefore not constant in (\ref{eq:A'Atilde}), but
varies with ${\vec x}'$; consequently,
(\ref{eq:A'Atilde}) does not define a macrostate of the system in
$\mathcal R'$.
In other words, the concept of macrostate is not Lorentz-invariant; indeed,
specifying the macrostate of a system in a given inertial frame does not
fix the macrostate for the same system in other inertial frames. From
the preceding
discussion, it should be clear that the same conclusion also applies to
microstates. Since the usual ensemble average is 
an average over all microstates
corresponding to a given macrostate, it is not therefore obvious 
that the procedure is Lorentz-invariant. To analyze further the situation, it
is necessary to introduce the new concepts of macro- and micro- histories.

Since the root of the apparent problem
lies in the fact that the concept of state is not Lorentz-invariant, the natural
idea is to replace that very concept by another one which is Lorentz-invariant.
Let us therefore introduce the concept of history and define the macrohistory
of a system by the values taken by various macroscopical fields at every
point in space-time where the system exists. In a given inertial frame
$\mathcal R$, this typically amounts to fixing the value of any of the
retained macroscopical fields at every point $\x$ in $\Rset^3$ for any 
value of $t$. The concept of microhistory will be defined accordingly.
It is clear from the discussion in the preceding paragraph that these
new concepts are Lorentz-invariant, \emph{i.e.}, specifying the macro- or
microhistory of a system in a given inertial frame is sufficient
to determine the macro- or micro- history of the same system in any other
inertial frame. If one then defines a relativistic statistical ensemble
$\Omega$ as the collection of systems with microhistories 
$\omega$ that correspond ---\emph{for a sufficiently short period of time
in the local co-moving system}--- to one and the same macrohistory,
the ensemble averaging procedure is, by construction, Lorentz-invariant and
the distribution function defined by (\ref{eq:deffgen}) is, therefore, indeed,
a Lorentz-scalar. Hence, we suppose that there exists collections of systems
that are microscopically different, but macroscopically identical for some
limited amount of time. This seems not too unrealistic an assumption.

Let us now prove that the covariant
ensemble average over histories gives back the results obtained by the usual 
average over states. In ordinary statistical mechanics, the macroscopical
fields always obey
deterministic equations and one can, in principle, given an inertial
reference frame $\mathcal R$, reconstruct the whole macrohistory of the system
from its macrostate at an arbitrary time $t_0$ in $\mathcal R$
through the evolution
equations. On the other hand, the microscopic degrees of freedom may obey
deterministic equations or stochastic equations.
If these equations are deterministic, the covariant
average over all microhistories
corresponding to a given macrohistory should coincide with the
usual ensemble average over all the microstates corresponding to the given
macrostate at time $t_0$ in $\mathcal R$. Indeed, through the
deterministic microscopic
evolution equations, one can then reconstruct the whole microhistories of the 
system from its microstates in $\mathcal R$ at an arbitrary time. In other 
words, deterministic evolution equations establish a one-to-one correspondence
between the histories and the states of the system at a given time in
$\mathcal R$. Thus averaging over histories comes down in such cases
to averaging over states.
The matter is  more complicated if the microscopic evolution
equations are stochastic, typically involving some random `noise'
(for an example 
of stochastic process in the relativistic framework, we refer the reader
to \cite{DMR97a} and \cite{BDR01a}).
For a given realization of the noise
\emph{i.e.} `freezing' the randomness, the stochastic evolution equations
act as deterministic ones and it is then possible to establish a one-to-one
correspondence between histories and states. Naturally, this correspondence
depends on the chosen realization. Keeping this realization fixed for the
moment, the average over histories again comes down to an
average over states.
To get a full ensemble average, one usually also averages
over the various realizations of the noise. Obviously, this final average
does not change the variance of the quantity to which it is applied.
Thus, the total ensemble average, including the average
over the realization of the noise, is a covariant operation in this case too.

To sum up these results\,: the notions of macro-and microhistories
are necessary to prove that the usual ensemble-averages over states
are indeed covariant operations. This result is not trivial because, 
 contrary to the concept of history, the notion of
 state is \emph{not} a covariant concept.

     \section{The concept of world-line distribution function}
     \label{sec:worldlines}

As we indicated in the introduction to this article, some authors introduce
\emph{ab initio} a new object in relativistic kinetic theory, the so-called
distribution-function for the world-lines of the particles and consider it
to be \emph{the} fundamental relevant concept for relativistic situations.
We would like, in this section, to prove that the existence of such a 
distribution is, somewhat counter-intuitively, \emph{not} trivial; we will
notably show that the world-line distribution, as it is usually defined
in the literature, exists only because the standard particle distribution
function is a Lorentz-scalar. In other words, to assume from the start that 
the world-line distribution function exists comes down to assuming
that the usual one-particle distribution in phase-space is a 
Lorentz-scalar.

      \subsection{Definition of the world-line distribution function}
      \label{ssec:defworldlinedist}

Let $\Sigma$ be any space-like hypersurface in
space-time and $d\Sigma^\mu$ its  normal four-vector at point $M(x^\mu$).
Let also $d^3p$ be a volume element in momentum space,
centered on a given three-momentum
$\vec p$. The world-line distribution function $f_w$ at point
$Q = (t, {\vec x}, \vec{p})$ is usually defined so that
the number of world-lines $dN_w$ 
with momentum in the range ($\vec{p}, d^3{\sl p}$) that cross
$d\Sigma^\mu$ in the direction of its normal is  given by\,:
\begin{equation}
dN_w = f_w (Q)\ p^\mu d\Sigma_\mu\ \frac{d^3p}{p^0}.
\label{eq:deffw}
\end{equation}
$p^0$ naturally is the time-component of the four-vector $p^\mu$ associated
to $\vec p$ by equation (\ref{23}).
Since $dN_w$, $p^{\mu} d\Sigma_{\mu}$ and ${d^3p}/{p^0}$ are
Lorentz-scalars, it follows from equation~(\ref{eq:deffw}) that $f_w$, 
if it exists, is also a Lorentz-scalar.
What makes definition~(\ref{eq:deffw}) not trivial is that the hyper-surface
$\Sigma$, aside from being space-like, is arbitrary. In particular, in any
given reference frame, one can choose to apply equation~(\ref{eq:deffw})
to an hyper-surface
which does \emph{not} coincide with the constant-time hyper-surfaces of this
frame. This is why $dN_w$ is not, in general, the  number of particles 
present in some infinitesimal volume of the phase-space at a given time in
the chosen reference frame, but a number of world-lines.

      \subsection{The reason why $f_w$ exists}
      \label{ssec:explfw}

Let $Q$ be any point in the one-particle phase-space, $M$ its projection
on the space-time manifold and $\mathcal R$, an
arbitrary inertial frame. It is always possible to find a space-like
hyper-surface $\Sigma$ which contains $M$ and the time-like 
normal vector of which, $d\Sigma$,   has vanishing space-components at $M$ in
$\mathcal R$\,:
\begin{equation}
d\Sigma_\mu = d^3x \delta^0_\mu.
\label{eq:dsigmamu}
\end{equation}
Transcribed  in this reference-frame, equation~(\ref{eq:deffw}) reads\,:
\begin{equation}
dW = f_w (t, {\vec x}, \vec{p}) d^3x d^3p.
\label{eq:dWusuel}
\end{equation}
Since the surface element $d\Sigma$ is, by construction, a constant-time
surface-element in $\mathcal R$, the number of world-lines $dW$
is also  the number of particles which, in this frame, occupy at time $t$ the 
phase-space volume $d^3x d^3p$. This proves that, if it exists, $f_w$ acts as
(and is) the standard
one-particle distribution function in $\mathcal R$ at point Q. 

Since $\mathcal R$ is arbitrary, this also proves that the definition of
$f_w$ only makes sense because the particle-distribution function
is frame-independent, \emph{i.e.}, because it is a Lorentz-scalar.
To phrase it slightly differently, if one has not yet proven that 
the one-particle distribution function is a Lorentz-scalar, assuming that
$f_w$ exists comes down to assuming that the particle distribution function is
a scalar. On the contrary, if one has proven (as was done in
section~(\ref{sec:macroh})) that the particle distribution
is a Lorentz-scalar, one can then introduce the invariant
world-line distribution,
prove that it exists and use it as a particularly elegant tool in 
manifestly covariant calculations. Let us also note that the very
concept of world-line distribution, as opposed to particle distribution, 
seems to imply the concept of statistical average over histories, as opposed to
average over states. All this is naturally perfectly coherent. Indeed, we have
just seen that the concept of world-line distribution function 
only makes sense because the particle distribution function is
a Lorentz-scalar, and it was proven in section {\ref{sec:macroh}}
that the most natural and general way to ensure that the particle
distribution is a Lorentz-scalar is precisely to use  covariant
statistical ensembles that are actually \emph{not} ensemble of states,
but ensembles of histories.

     \section{Discussion}
     \label{sec:Discu}

In this article, we have given a fresh look at the notion of relativistic 
distribution function commonly used in relativistic kinetic theory. Let us
sum up our main results. As already assumed by various authors, the standard
Galilean definition for the one-particle distribution function in phase-space
can be imported safely into the special-relativistic realm. It is then
possible to prove that this distribution is a Lorentz-scalar. However, the two
direct proofs that exist in the literature have been found wanting. The first
one is  based on the so-called invariance of the volume-element in one-particle
phase-space; we have proven by direct calculation that, contrary to earlier
claims made by various authors, this volume-element
is \emph{not} Lorentz-invariant, and the whole proof has been shown
to rest on a misconception of the problem.
The second proof, due to de~Groot and
Suttorp (and later on incorporated in de~Groot's, van~Leeuwen's and van~Weert's
book on relativistic kinetic theory), starts from the
most general definition of the one-particle distribution-function. We have
shown that, to be fully convincing, this proof needs the introduction of the
new manifestly-covariant (relativistic) concepts of micro- and macrohistories.
With these notions, new, covariant statistical ensembles can be introduced and 
the one-particle distribution function can be shown to be indeed a
Lorentz-scalar.
We have also revisited the axiomatic approach to relativistic kinetic theory,
which starts by introducing the non-Galilean concept of distribution 
function for the world-lines crossing an arbitrary space-like 
hypersurface in space-time. We have shown that
introducing this new concept \emph{ab initio} in the relativistic theory
is tantamount to assuming axiomatically that the usual particle
distribution in phase-space is frame-independent. Since this is a fact
which can be proven, as can be clearly seen from the argument of the 
present article, it seems to us that assuming it from scratch is unnecessary.
On the other hand, building on the scalar-nature of the 
one-particle distribution to construct the distribution
function for world-lines is certainly interesting, since the world-line 
distribution function is a most useful tool in manifestly
covariant calculations.   
 
This article would not be complete without a mention of another proof 
that the relativistic one-particle distribution in phase-space is a 
Lorentz-scalar. This proof \cite {BDR01a} is actually
 rather particular because it has been
given in the context of relativistic stochastic processes only, and more 
precisely for the distribution function associated to the so-called
relativistic Ornstein-Uhlenbeck process, which is a toy-model of
relativistic diffusion. As such, this proof makes extensive use of 
stochastic calculus. How this connects with the general proof envisaged
in this article is not absolutely clear yet and we hope to shed 
further light on this question in a forthcoming publication.

As also mentioned in the introduction, this article tackled with
the special relativistic situation only. The general relativistic case
is addressed in the article companion to the present one;
envisaging the problem in an arbitrary reference-frame naturally contributes
to a deeper understanding of the simpler, special-relativistic case, where
the discussion is restricted to inertial frames only.
     
\par\noindent
ACKNOWLEDGMENTS\,: The authors wish to acknowledge fruitful
discussions with C\'ecile Barbachoux.


\begin{thebibliography}{99}

\bibitem{LL80a}
  L.D.~Landau and E.M.~Lifschitz, \emph{Statistical Physics},
  Vol.~5, 3rd edition, Pergamon Press (1980).

\bibitem{H63a}
  K.~Huang, \emph{Statistical Mechanics}, John Wiley and Sons (1963).

\bibitem{E71a}
  J.~Ehlers,  \emph{General Relativity and Kinetic Theory} in
  \emph{General Relativity and Cosmology}, R.K.~Sachs Ed.,
  Academic Press (1971).

\bibitem{I89a}
  W.~Israel, in \emph{Relativistic Fluid Dynamics}, eds. A.~Anile and
  Y.~Choquet-Bruhat (Springer-Verlag, Berlin 1989).

\bibitem{GLW80a}
  S.R.~de Groot, W.A.~van Leeuwen and Ch.G.~van Weert, \emph{Relativistic
  Kinetic Theory} (North Holland, Amsterdam 1980).

\bibitem{LL75a}
  L.D.~Landau and E.M.~Lifschitz, \emph{The Classical Theory of Fields},
  Vol.~2, 4th edition, Pergamon Press (1975).

\bibitem{MTW73a}
  C.W.~Misner, K.S.~Thorn and J.A.~Wheeler, \emph{Gravitation}
  (W.H.~Freeman and Co., New-York 1973)

\bibitem{GS72a}
  S.R.~de~Groot, L.G.~Suttorp, \emph{Foundations of Electrodynamics}
  (North-Holland Publishing Company, Amsterdam, 1972).

\bibitem{DMR97a}
  F.~Debbasch, K.~Mallick and J.P.~Rivet, 
  \emph{J.~Stat.~Phys.} {\bf 88}(3/4), p.~945, 1997.

\bibitem{BDR01a}
  C.~Barbachoux, F.~Debbasch and J.P.~Rivet, \emph{The spatially
  one-dimensional relativistic Ornstein-Uhlenbeck process in an
  arbitrary inertial frame},
  \emph{Eur. Phys. J.} B, {\bf 19}, pp.~37--47, 2001.

\end{thebibliography}
\end{document}